\documentclass[11pt]{article}
\usepackage{cospar}
\usepackage{url}

\usepackage{graphicx}
\usepackage[figuresright]{rotating}

\title{HST and VLT observations of Isolated Neutron Stars}

\author{R. P. Mignani\address{European Southern Observatory,
Karl-Schwarzschild-Str.\ 2, D85740, Garching, Germany},
 G. G. Pavlov\address{Pennsylvania State University, 525 Davey Lab, University Park, PA 16802, USA}, 
A. De Luca\address{Istituto di Astrofisica Spaziale e Fisica Cosmica, Sezione di
Milano "G.Occhialini" - CNR v. Bassini 15, I-20133 Milan, Italy} and 
P. A. Caraveo\address{Istituto di Astrofisica Spaziale e Fisica Cosmica, Sezione di
Milano "G.Occhialini" - CNR v. Bassini 15, I-20133 Milan, Italy}
}
\begin{document}

\maketitle

\begin{abstract}
New results of HST and  VLT observations of isolated pulsars and their
environments   are  presented.We  present   the  first   deep  optical
observations of  the nearby pulsar PSR  J0108-1431, the identification
evidence of PSR 1929+10 based on the proper motion measurement of its
optical counterpart and a deep investigation at optical wavelengths of
the X-ray PWN around Vela.
\end{abstract}


\section*{PSR J0108-1431}

PSR J0108$-$1431  was discovered by  Tauris et al.\ (1994)  during the
Parkes Southern Pulsar Survey.  With a period $P=0.808$ s and a period
derivative  $\dot{P} =  7.44 \times  10^{-17}$ s~s$^{-1}$  (D'Amico et
al.\ 1998),  the pulsar  has a characteristic  age $P/2\dot{P}  = 170$
Myr,  rotation  energy   loss  rate  $\dot{E}=5.6\times  10^{30}$  erg
s$^{-1}$, and magnetic field $B=2.5\times 10^{11}$ G.  From the Taylor
\& Cordes  model (1993), the  low dispersion measure ($2.38  \pm 0.01$
cm$^{-3}$~pc)  puts  PSR J0108$-$1431  at  a  distance  of 60-130  pc.
Although very  close, PSR  J0108-1431 has never  been detected  so far
outside the radio band.   Virtually undetectable according to standard
cooling models, which predict a temperature $T < 10^4$ K for an age of
$\sim$  170 Myr (Tsuruta  1998), the  detection of  optical-UV thermal
radiation  from the  neutron  star  (NS) surface  would  allow one  to
constrain possible  heating mechanisms like the  dissipation of energy
of  differential rotation  (e.g.   Van Riper  et  al.~1994) and  Joule
heating caused  by dissipation of the  magnetic field in  the NS crust
(Miralles, Urpin  \& Konenkov 1998). Thus, if  some heating mechanisms
indeed operate,  one can  expect the surface  NS temperature of  a few
times  $10^4$~K,  detectable in  the  optical-UV  but undetectable  in
X-rays.    Optical observations of  PSR J0108-1431 were  carried out
with the 6-m telescope  of Special Astrophysical Observatory (Russia),
but  they were  not deep  enough to  detect the  pulsar (Kurt  et al.\
2000).  Therefore, we performed a  deep observation with the Antu unit
of  the  ESO Very  Large  Telescopes  (VLT)  between July  and  August
2000. Images  were obtained  in the  Bessel filters  $U$  (9000s), $B$
(5400s), and  $V$ (7200s) using  the FOcal Reducer and  Spectrograph 1
(FORS1)  instrument operated  at  its standard  angular resolution  of
0.2'' per  pixel, with a corresponding  field of view  of $6.8' \times
6.8'$.  See Mignani,  Manchester and Pavlov (2003) for  details on the
observations and the data reduction.  To obtain an updated reference
radio position of  the pulsar, we performed new  observations with the
Australia Telescope Compact Array (ATCA) on 2001 March 31 in two bands
centered on 1384 MHz and 2496  MHz.  We derived a mean position (epoch
2001.3) of       $\alpha_{J2000}$=$01~08~08.317      \pm      0.010s$,
$\delta_{J2000}$=$-14~31~49.35  \pm   0.35"$,  coincident  within  the
uncertainties with that obtained from the timing data (D'Amico et al.\
1998).  This  implies an  upper limit on  the pulsar proper  motion of
82~mas~yr$^{-1}$, corresponding  to a total velocity $v  < 50 d_{130}$
km s$^{-1}$ (with $d_{130}$ being  the pulsar distance in units of 130
pc)  i.e.  within  the bottom  10\%  of the  measured pulsar  velocity
distribution  (Lorimer, Bailes  and  Harrison 1997).   By  using as  a
reference the  positions of stars selected  from the GSC  II, the most
recent  pulsar position  was finally  registered on  the  FORS1 images
(overall accuracy $0.33''$  in R.A.  and $0.46''$ in  Dec). The pulsar
position at the  epoch of our optical observations  is marked in Fig.1
(left).  To facilitate the object  detection, all the images have been
smoothed  using a gaussian  filter.  The  pulsar position  falls about
$\approx 0.6''$  East from an  elliptical galaxy in the  field.  Since
such  a distance  corresponds to  about twice  the uncertainty  of the
pulsar  position in  R.A, it  is unlikely  that the  pulsar  is hidden
behind the  galaxy, but  we cannot rule  it out.  Only  few point-like
objects (labelled in  Fig.1, left) have been detected  within a radius
of $6''$  from the pulsar position  but they are too  distant from the
nominal    pulsar     position    to    be     considered    candidate
counterparts. Therefore,  we conclude  that no optical  counterpart to
the pulsar  can be  identified in  our data down  to $3  \sigma$ upper
limits of $V \simeq 28$, $B\simeq 28.6$ and $U \simeq 26.4$ e.g.  more
than  3  magnitudes  deeper  than  those  obtained  by  Kurt  et  al.\
(2000).   For  a distance  of 130  pc, estimated  from  the pulsar's
dispersion measure, our  constraints on the optical flux  put an upper
limit  of $T=4.5\times  10^4$ K  for  the surface  temperature of  the
neutron star, assuming a stellar radius $R_\infty=13$ km.

 \begin{figure}            
\includegraphics[width=75mm]{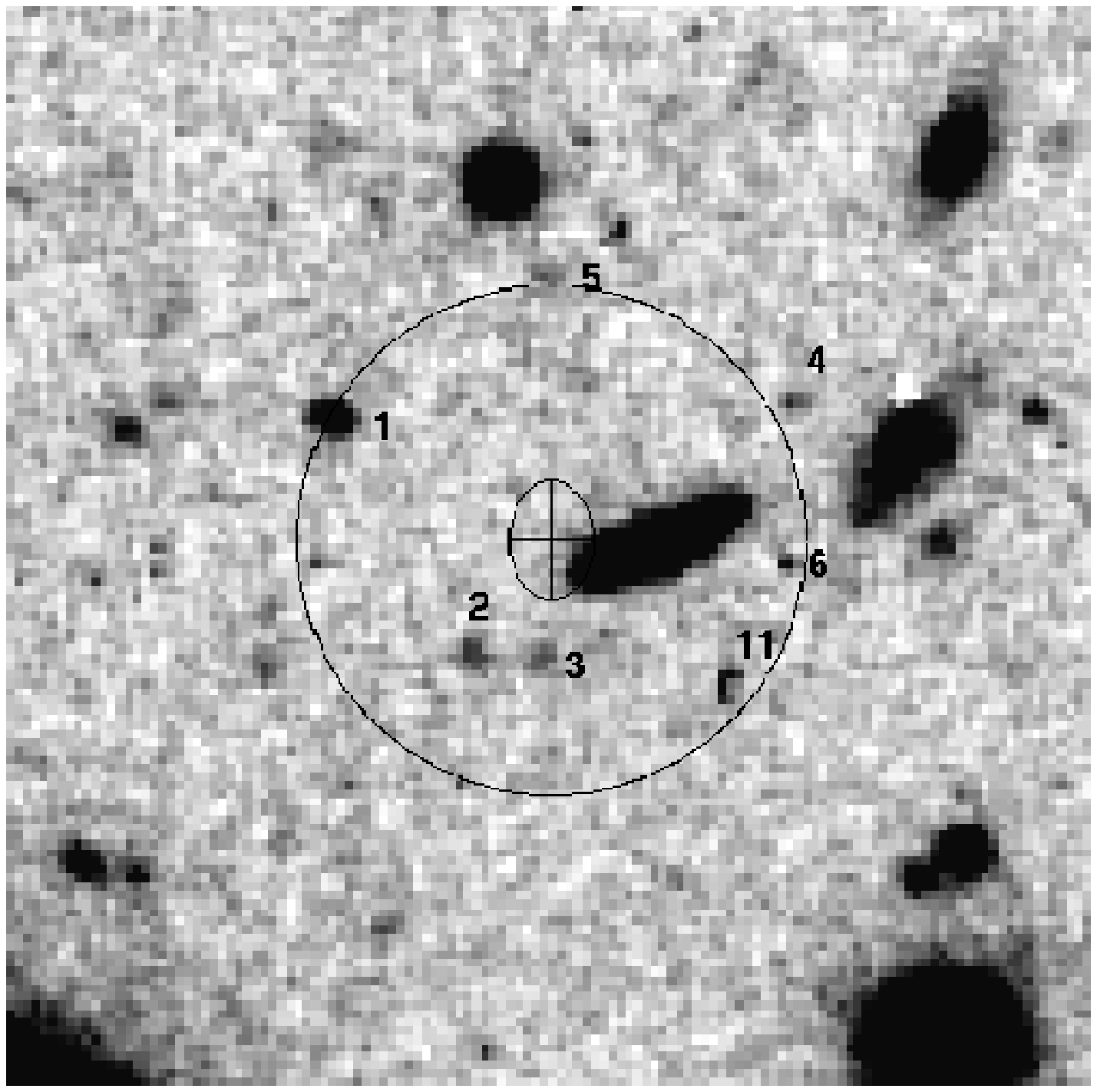}
\includegraphics[width=65mm]{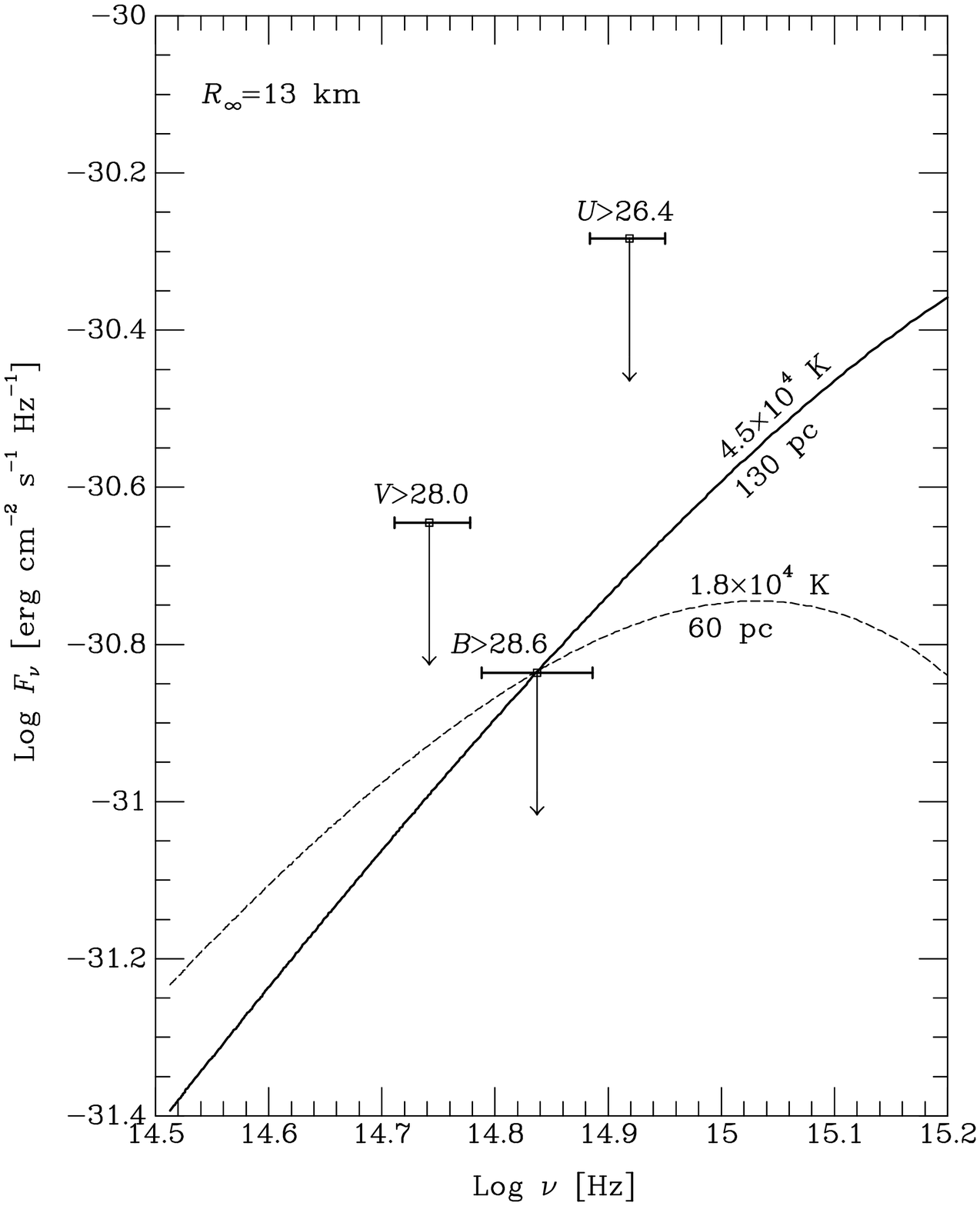}   
\caption{(left) FORS1 $25'' \times 25''$ $V$-band image (7200s) of the
PSR  J0108$-$1431 field (North  to the  top, East  to the  left).  The
cross marks  the nominal radio position  of the pulsar,  with the arms
equal  to 3  times the  overall  uncertainty on  the pulsar  position.
Point-like objects detected at $\ge 3 \sigma$ in at least one passband
and located  within (close to) a  $\sim 6''$ radius  around the pulsar
position are labelled.   (right) Upper limits on the  pulsar fluxes in
the  $UBV$  bands and  blackbody  spectra  corresponding  to the  most
stringent $B$-band  limit for  the distances  of 60 and  130 pc  and a
neutron star radius $R_\infty = 13$ km.}  \end{figure}

\section*{PSR B1929+10}

PSR1929+10  is also  an old  ($ \sim  3~ 10^{6}$  yrs)  and relatively
nearby  ($\sim$  330  pc  -   Brisken  et  al.   2002)  radio  pulsar.
Originally detected  in X-with Einstein (Helfand  1983), pulsations at
the radio  period (227  ms) were discovered  by ROSAT  (Yancopulous et
al. 1994)  and confirmed by ASCA  (Wang and Halpern  1997).  The X-ray
spectrum  could  be either  thermal  ($T  \sim  3-5 ~  10^{6}~K$)  and
produced  from hot  polar caps  (Yancopulous  et al.   1994; Wang  and
Halpern  1997) or non-thermal  with a  spectral index  $\alpha \approx
1.27  \pm  0.4$ (Becker  and  Tr\"umper  1997).   A candidate  optical
counterpart was identified  with the HST/FOC by Pavlov  et al.  (1996)
from the positional coincidence with the radio coordinates.  To secure
the  identification of the  PSR B1929+10  optical counterpart  we have
applied the strategy  used by Mignani et al.(2000),  i.e., the measure
of a proper motion consistent with  the radio one. The field of PSR
B1929+10 was imaged with the HST/STIS between August and October 2001.
Ten 1200s exposures  were obtained over five visits  with the NUV-MAMA
detector ($24.7'  \times 24.7'$ field  of view, $0.024''$ pixel  size) through the
F25QTZ filter ($\lambda=2364 \AA, \Delta \lambda \sim 842 ~\AA~FWHM$).
The STIS time-integrated images were calibrated using the standard HST
pipeline and corrected for  the CCD geometric distortion.  As starting
point  for our  proper motion  measurement  we used  the  FOC
observations of Pavlov  et al.  (1996), retrieved from
the ST-ECF public archive and recalibrated on-the-fly.    A complete
description  of  the relative astrometry procedure  with  a  detailed
discussion of  the error budget  is reported in Mignani  et al.(2002).
After  the registration  of the  different  epoch images  in a  unique
reference frame,  we could  measure the candidate  optical counterpart
displacement over  the 7.2 years (Fig.2, left).   The five independent
measurements yielded  results fully consistent within  the errors.  By
means of  a simple $\chi^{2}$ fit  we obtained the  best proper motion
values:   $\mu_{\alpha}cos(\delta)   =   +97  \pm   1$   mas~yr$^{-1}$
$\mu_{\delta} =  +46 \pm 1$ mas~yr$^{-1}$ corresponding  to a position
angle  (PA)  of  64.63$^{\circ}$  $\pm$  0.55$^{\circ}$.   Within  the
errors,  these results  are  fully compatible  in  both magnitude  and
direction   with   the   radio    ones   (Brisken   et   al.    2002):
$\mu_{\alpha}cos(\delta)= +94.82 \pm 0.26$ mas~yr$^{-1}$ $\mu_{\delta}
=   +43.04   \pm   0.15$   mas~yr$^{-1}$   (PA=65.58$^{\circ}$   $\pm$
0.09$^{\circ}$).  Our proper motion  measurement thus provide a robust
proof that the candidate proposed by Pavlov et al.(1996) is indeed the
optical counterpart of  PSR B1929+10.   The STIS/F25QTZ  flux of the
pulsar has been compared with  the photometry of Pavlov et al.  (1996)
in  the   F130LP  ($\lambda=3437.7  \AA,  \Delta   \lambda  \sim  1965
~\AA~FWHM$)  and F342W  ($\lambda=3402  \AA, \Delta  \lambda \sim  442
~\AA~FWHM$) filters.   At a variance  with middle-aged ($\sim$  $10^{5}$ yrs)  pulsars such  as  PSR B0656+14,  PSR B1055-52  and
Geminga, for which the optical fluxes are somewhat compatible with the
extrapolations of  their X-ray  spectra, the flux  of the  PSR 1929+10
candidate counterpart deviates by about 3 orders of magnitude from the
predicted values  (Fig.2, right).  In  addition, the data seems  to be
consistent with a power law, with  spectral index $\alpha = 0 \div 1$,
depending  on the  actual  value  of the  reddening.   This result  is
apparently  in  contrast  with   the  available  data  from  young  to
middle-aged pulsars  which suggest, as  a general trend,  a decreasing
importance of non-thermal processes with the age.

\begin{figure}
 \includegraphics[width=75mm]{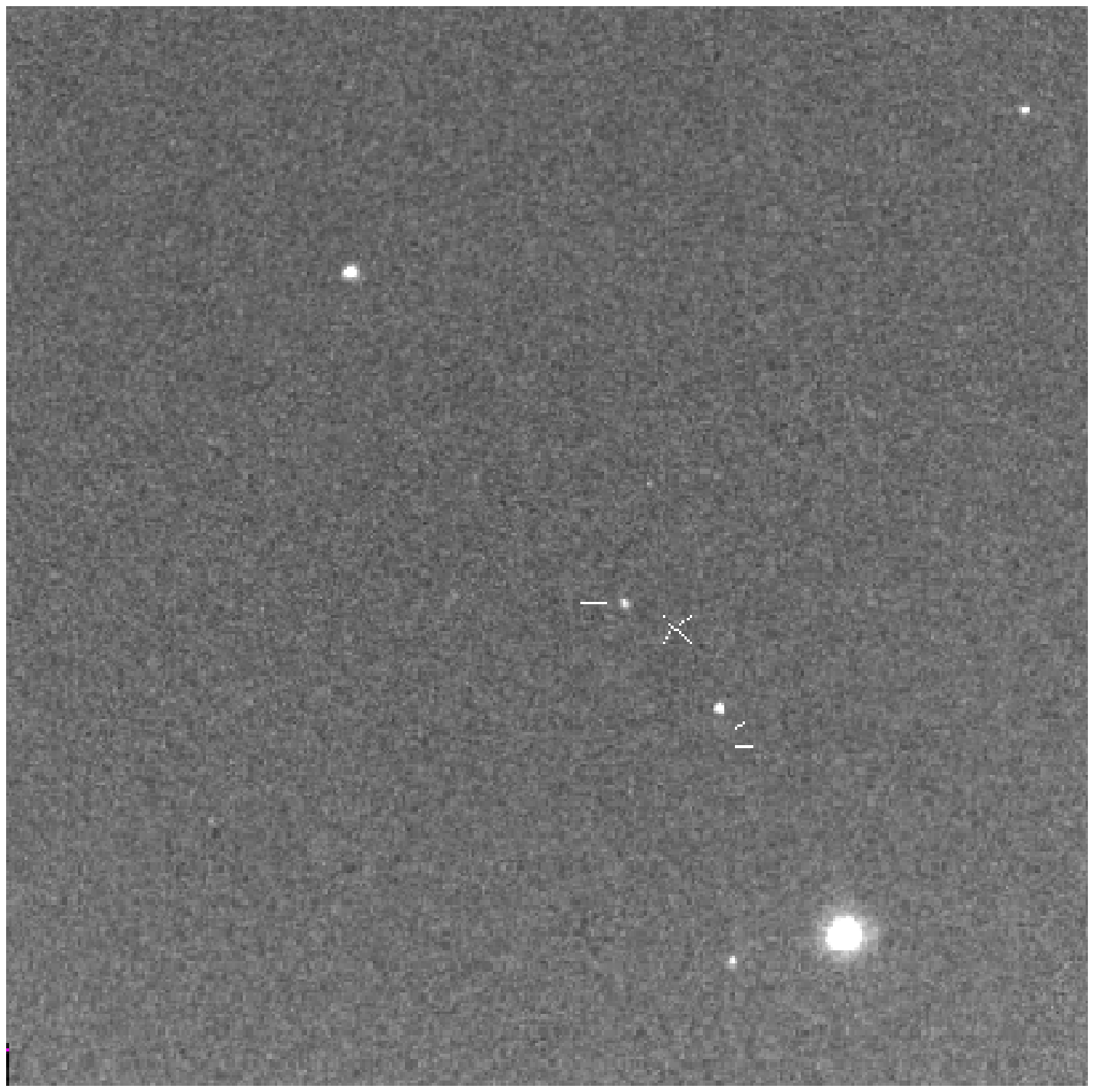}
 \includegraphics[width=75mm,clip]{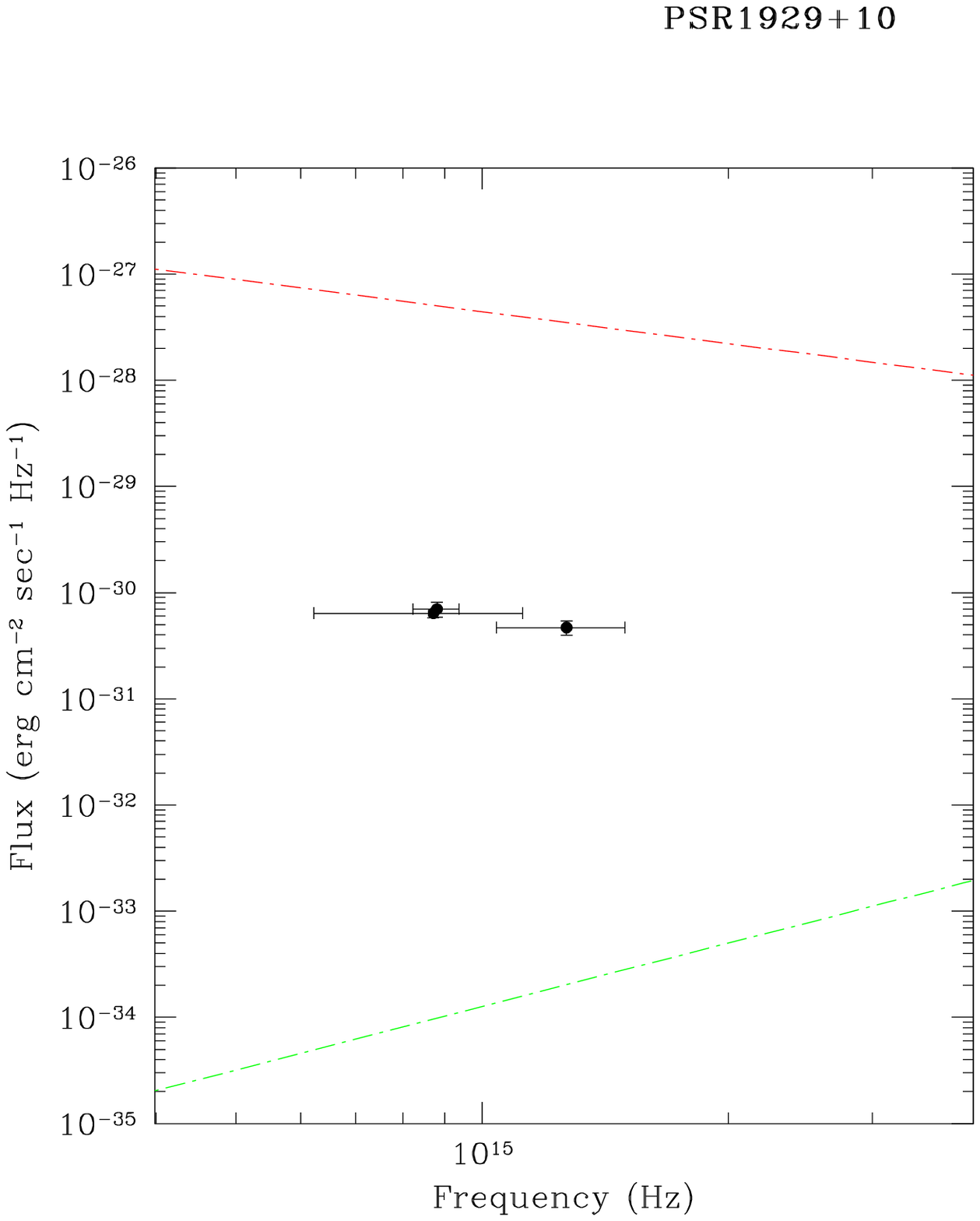}     
\caption{(left)
 STIS/NUV-MAMA/F25QTZ image  of the PSR 1929+10  field (12000S). North
 to the  top, East to the  left.  The pulsar  candidate counterpart is
 marked by the two ticks.  The  cross marks the position of the pulsar
 at epoch 1994.52. (right)  Multiband optical flux distribution of PSR
 1929+10 from the  photometry of Mignani et al.  (2002) and (Pavlov et
 al. 1996).  The two dashed lines represents the optical extrapolation
 of the best fitting X-ray spectral models. } \end{figure}

\section*{The Vela pulsar optical nebula}

A compact ($\sim 4'$) X-ray nebula around the Vela pulsar was detected
 in soft X-rays with the {\sl Einstein} HRI (Harnden  et al. 1985) and
 firstly  interpreted  as   a  pulsar-wind   nebula (PWN) powered   by
 relativistic  particles ejected by  the  pulsar, which  shock  in the
 ambient medium    and  emit synchrotron    radiation.   Based on  its
 ``kidney-bean''  shape,  Markwardt  \& \"Ogelman   (1998) proposed an
 interpretation in terms   of a bow-shock  produced  by the supersonic
 motion of the pulsar through the ambient medium.  The Vela pulsar and
 its PWN have  been recently observed  with  both the High  Resolution
 Camera  (HRC)  and the Advanced  CCD   Imaging Spectrometer (ACIS) on
 board  the {\sl  Chandra} X-ray observatory.  Thanks to the  excellent angular resolution  of
 {\sl Chandra},  the morphology of the  PWN was resolved  in a complex
 structure, similar to that of the Crab PWN, which cannot be explained
 by a simple bow-shock model.  The  brighter part of the PWN ($\approx
 2'$), shows an approximately axisymmetric structure, with two arcs, a
 jet,  and a counter-jet, embedded  into an extended diffuse emission.
 The axis of   symmetry,  which can  be  associated  with  the  pulsar
 rotational axis, nearly coincides with the  direction of the pulsar's
 proper motion (P.A.  = 301$^{\circ}$ --- e.g.,  Caraveo et al.\ 2001a).
 The inner PWN is   then surrounded by   a bean-shaped diffuse  nebula
 ($\sim  2'\times 2'$) with an  elongated   region of fainter  diffuse
 emission southwest of the brighter part of the PWN and detected up to
 $\sim 4'$ from  the pulsar.  A $100''$-long  outer  jet is also  seen
 northwest of the  inner PWN.  The overall  spectrum of the PWN can be
 described  by  a power law  with an   average spectral (energy) index
 $\alpha \approx 0.5$,
which
 can be interpreted as  synchrotron emission of relativistic electrons
 and/or positrons.  Such  a spectrum is expected  to extend to optical
 and radio frequencies.  Although the radio emission from the compact nebula
 is  probably  dominated by  the   brightness  of the   pulsar, highly
 polarized (up to  60\% at 5 GHz)  extended ($\sim\!4'$) radio emission
 has been   indeed detected in the surrounding   region (Lewis et al.\
 2002;  R.\ Dodson 2002, private communication)   with two radio lobes
 southeast ($\sim\!~18$     arcmin$^2$)   and  northeast    ($\sim\!5$
 arcmin$^2$) wrt  the pulsar.  One can  then expect the presence of an
 optical nebula  around the Vela pulsar,  as observed in several other
 young pulsars like the Crab PWN (see, e.g., Hester  et al.\ 2002, and
 references   therein) and PSR    B0540--69  (Caraveo et  al.\  2001b). Contrary to the  Crab and PSR  0540-69, searches for the  Vela
 PWN  in optical have  been  inconclusive so  far.  \"Ogelman et  al.\
 (1989) reported only a marginal detection of a putative optical PWN ($\sim
 2'$).

\begin{table*}[h]
\begin{center}
\begin{tabular}{lccccclcc} \hline
{\rm Tel.} & {\rm Instr.} & {\rm FOV}  & {\rm Scale} & {\rm Date}  &{\rm Filter} & $\lambda~(\Delta \lambda)$ & {\rm  Exp. (s)}  \\ \hline
NTT & EMMI-B & $6.2'\times 6.2'$ & 0.37'' & Jan 1995 & $U$    & 3542\AA\ (542\AA)  & 4800   \\
    & EMMI-B & $6.2'\times 6.2'$ & 0.37'' & Jan 1995 & $B$    & 4223\AA\ (941\AA)  & 1800  \\
    & EMMI-R & $9.2'\times 8.6'$ & 0.27'' & Jan 1995 & $V$    & 5426\AA\ (1044\AA) & 1200   \\
    & EMMI-R & $9.2'\times 8.6'$ & 0.27'' & Jan 1995 & $R$    & 6410\AA\ (1540\AA) & 900    \\
\hline
{\sl HST} & WFPC2  & $2.6'\times 2.6'$ & 0.1'' & Jun 1997 & 555W & 5500\AA\ (1200\AA)  & 2600   \\
          & WFPC2  & $2.6'\times 2.6'$ & 0.1'' & Jan 1998 & 555W & -           & 2000  &  \\
          & WFPC2  & $2.6'\times 2.6'$ & 0.1'' & Jun 1999 & 555W & -           & 2000  & \\
          & WFPC2  & $2.6'\times 2.6'$ & 0.1'' & Jan 2000 & 555W & -           & 2600  &  \\
          & WFPC2  & $2.6'\times 2.6'$ & 0.1'' & Jul 2000 & 555W & -           & 2600  &  \\
          & WFPC2  & $2.6'\times 2.6'$ & 0.1'' & Mar 2000 & 675W & 6717\AA\ (1536\AA) & 2600   \\
          & WFPC2  & $2.6'\times 2.6'$ & 0.1'' & Mar 2000 & 814W & 7995\AA\ (1292\AA) & 2600    \\
\hline
{\sl VLT} & FORS1 & $6.8'\times 6.8'$ & 0.2'' & Apr 1999 & $R$    & 6750\AA\ (1500\AA) & 300    \\
          & FORS1 & $6.8'\times 6.8'$ & 0.2'' & Apr 1999 & $I$    & 7680\AA\ (1380\AA) & 300    \\
\hline 
ESO/MPG 2.2m & WFI & $34'\times 33'$ & 0.24'' & Apr 1999 & H$_{\alpha}$ & 6588\AA\
(74.3\AA) & 3600      \\ \hline
\end{tabular}
\end{center}
\caption{Available optical datasets for the Vela pulsar
field. The first four columns list the telescope and the detector used
for  the  observations,  the  field  of  view  and  the  pixel  scale,
respectively. The epoch  of observation is in column  five. The filter
names  are listed  in column  six,  with their  pivot wavelengths  and
widths in column seven. The total integration time per observation (in
seconds) is given in column eight. }
\end{table*}

\begin{figure}[t]
\begin{center}
 \includegraphics[width=83mm]{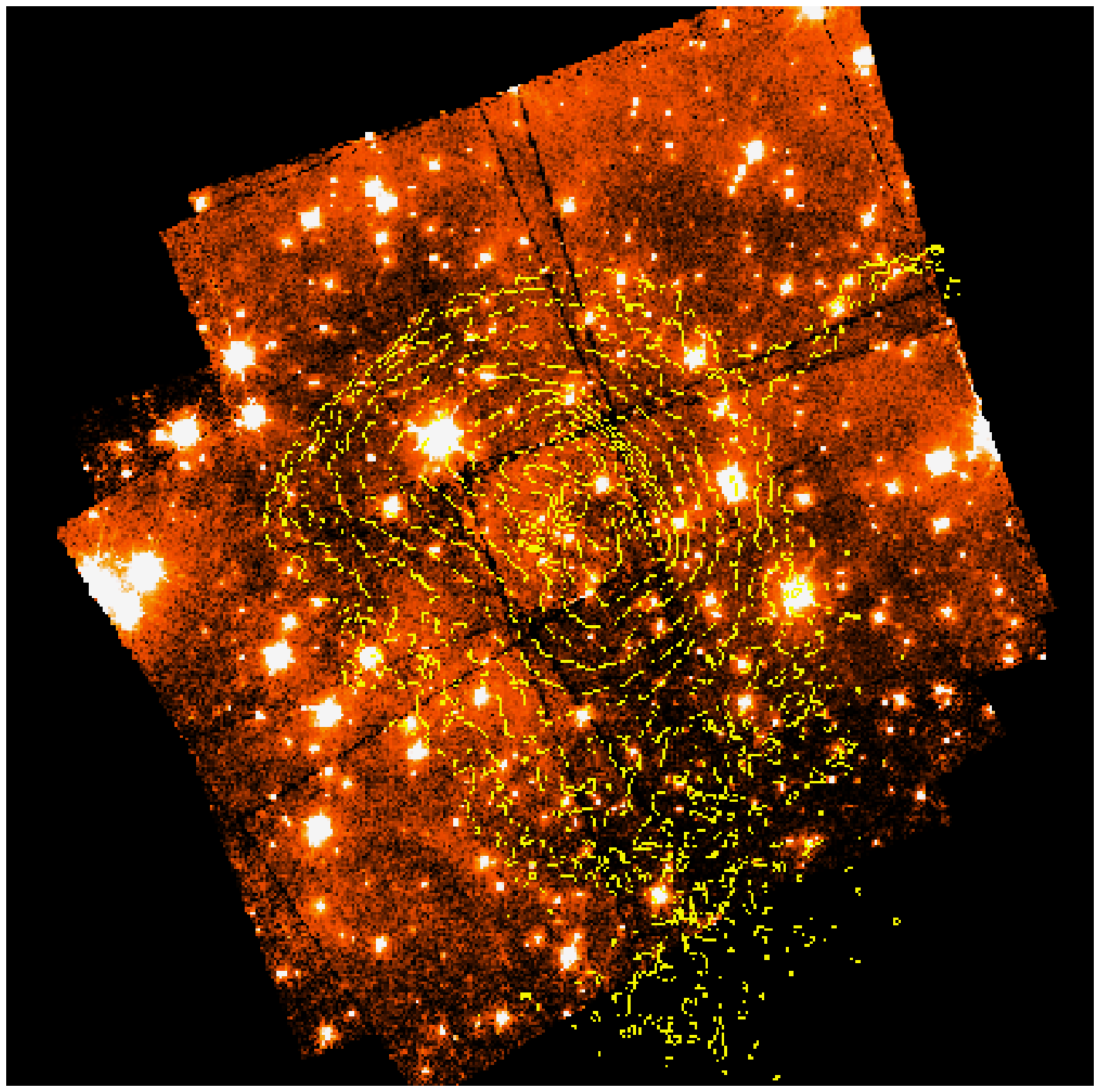}
 \includegraphics[width=81mm]{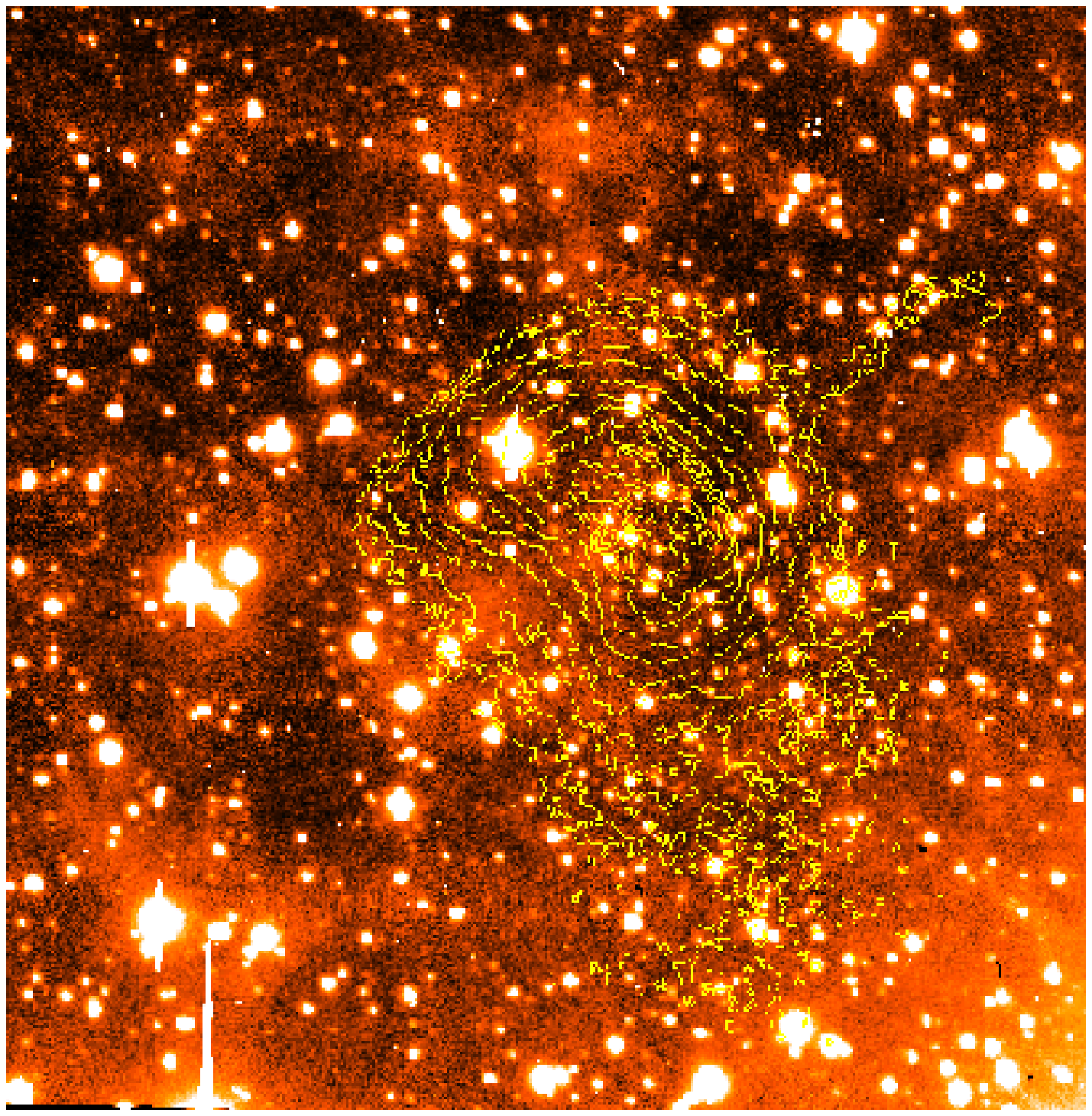} 
\end{center}
\caption{(left)  Combined WFPC2/555W  image of  the Vela  pulsar field
(11800s). North to the top,  East to the left.     The pulsar position  is located  within the
innermost  contour. (right)  Combined $UBVR$ image of  the Vela
pulsar field  obtained from the {\sl NTT}/EMMI observations.
The overlayed contours
(logarithmic scale)  correspond to  the X-ray intensity  maps obtained
from the  {\sl Chandra}  ACIS image of  the field integrated  over the
energy band  (1--8 keV).}
\end{figure}

\noindent
Here we  discuss the   results  of spatial correlations    between the
 recently obtained   {\sl Chandra}  ACIS images   and  {\sl HST} WFPC2
 images,  together with  observations  obtained   with the ESO    {\sl
 NTT}/{\sl VLT}/{\sl 2.2m} telescopes (Mignani et al. 2003). See Table
 1 for a detailed summary of the optical observations.  For a direct
 compararison,  optical and  X-ray images have  been superimposed with
 respect to the  absolute ($\alpha$,$\delta$) reference frame, relying
 on the  astrometric  solution  of  each image.  The  absolute   frame
 registration  between the optical  and X-ray images  turned out to be
 accurate   within   {$\approx   1''$--$2''$},   comparable  with  the
 uncertainty in the absolute astrometry  of each frame.  Our  starting
 point was the combined WFPC2 F555W image, which is by far the deepest
 optical image of the Vela pulsar field.  The  final image is shown in
 Fig. 3  (left), where  we have superimposed    the X-ray contour  map
 obtained from the combined {\sl Chandra} ACIS exposure of the region.
 The point  source within the innermost  X-ray  contour is the optical
 counterpart of the Vela pulsar.  The central PC field is large enough
 to map the inner part of the X-ray PWN,  i.e., the inner arc, the jet
 and the    counter-jet,  easily  identifiable in    the contour  map.
 Unfortunately, the outer arc region  is partially coincident with the
 CCD gaps between the PC and WFC chips.   No optical counterparts of
 the known X-ray features can be identified in the combined WFPC2/555W
 image, nor any other structure symmetric to  the pulsar proper motion
 direction (Fig.3 - left).  We set 3$\sigma$ upper  limits of 27.4 mag
 arcsec$^{-2}$  (0.39  $\times   10^{-30}$  ergs  cm$^{-2}$   s$^{-1}$
 Hz$^{-1}$ arcsec$^{-2}$)  and 27.1  mag  arcsec$^{-2}$ (0.57  $\times
 10^{-30}$ ergs  cm$^{-2}$ s$^{-1}$  Hz$^{-1}$ arcsec$^{-2}$)  for the
 innermost region (inner arc, jet, and counter-jet) and the outer arc,
 respectively.       After  correcting for  interstellar   extinction,    these  values translate  to
 $\approx  27.0$ mag   arcsec$^{-2}$   (0.52 $\times 10^{-30}$    ergs
 cm$^{-2}$ s$^{-1}$   Hz$^{-1}$ arcsec$^{-2}$) and  $\approx 26.7$ mag
 arcsec$^{-2}$  (0.57 $\times   10^{-30}$   ergs   cm$^{-2}$  s$^{-1}$
 Hz$^{-1}$ arcsec$^{-2}$), respectively.   The  analysis  was repeated
 for the  rest of  the  optical database, but    no evidence for  a
 compact optical nebula was found.  We used the wider {\sl NTT} images
 to search for diffuse optical emission up to distances of $\approx 3$
 arcmin, in particular at  the position of  the southwest extension of
 the  X-ray nebula.   Although   many background enhancements   can be
 recognized in  the combined {\sl NTT} $UBVR$   image (Fig.3 - right),
 they  can  be also    identified  in the   ESO/2.2m   H$_\alpha$ one,
 suggesting  that they are most    likely associated with the   bright
 filaments of the Vela supernova remnant. We thus  put upper limits on
 the  extended emission   of   25.4  mag arcsec$^{-2}$  (2.5   $\times
 10^{-30}$  ergs cm$^{-2}$  s$^{-1}$ Hz$^{-1}$  arcsec$^{-2}$).  We
 have compared the measured optical upper  limits on the brightness of
 the Vela PWN for the outer arc and the  diffuse emission southwest of
 the pulsar with the extrapolation of the X-ray  and radio data. While
 the optical upper  limits for the inner/outer  arcs are close to such
 extrapolation, in the case of the southwest diffuse emission they lie
 about  3 orders of  magnitudes  above the expected  value (Mignani et
 al. 2003). To summarize, we conclude that it is very likely that just
 a slightly deeper optical observation  would allow one to detect  the
 outer/inner arc (and other  bright elements of  the inner PWN), while
 even deeper optical observations are   needed to detect the  emission
 from the Vela PWN at large.

%

\end{document}